\documentclass[a4paper,12pt]{article}
\usepackage[utf8x]{inputenc}
\usepackage{amsfonts}
\usepackage{amsmath}
\usepackage{amssymb}
\usepackage{amsthm}
\usepackage{booktabs}
\usepackage{float}
\usepackage{multirow}
\usepackage{graphicx} 
\usepackage{graphics}
\usepackage[round]{natbib}
\usepackage{url}
\usepackage{hyperref}
\usepackage{longtable}
\usepackage[]{threeparttable}
\usepackage{rotating}
\usepackage{xcolor}
\usepackage{comment}

\newcounter{example}

\title{Estimation of Dirichlet distribution parameters with bias-reducing adjusted score  functions}
\author{V. GIOIA$\,^1$ and E. C. KENNE PAGUI$\,^2$\\
$\,^1\,$\small University of Udine, Department of Economics and Statistics \\
$\,^2\,$\small University of Padova, Department of Statistical Sciences\\
\small gioia.vincenzo@spes.uniud.it, eulogeclovis.kennepagui@unipd.it
}
\date{}
\begin{document}

\maketitle

\begin{abstract}
\noindent
The Dirichlet distribution, also known as multivariate beta, is the most used to analyse frequencies or proportions data. Maximum likelihood is widespread for estimation of Dirichlet's parameters. However, for small sample sizes, the maximum likelihood estimator may shows a significant bias. In this paper,  Dirchlet's parameters estimation is obtained through modified score functions aiming at mean and median bias reduction of the maximum likelihood estimator, respectively. A simulation study and an application compare the adjusted score approaches with maximum likelihood.

\noindent

\end{abstract}

\noindent
\emph{Some key words:} Compositional data,   likelihood,  bias reduction.

%%%%%%%%%%%%%%%%%%%%%%%%%%%%%%
\section{Introduction}
%%%%%%%%%%%%%%%%%%%%%%%%%%%%%%
Proportions data, also referred as compositional data, are very pervasive in many disciplines, ranging from natural sciences to economics. Dirichlet distribution, that is a multivariate generalization of the beta distribution and belongs to the exponential family, is the simplest choice to handle with proportions.  Inference on parameters is easily carried out with maximum likelihood (ML). However, for small sample size and large number of parameters, the ML estimator exhibits a relevant bias, as is apparent in simulation results of \citet{Narayanan}. 

This paper aims to improve the ML estimates by using modified score functions. Following \citet{Firth}, the mean bias reduced (mean BR) estimator is obtained as solution of a suitable modified score equation. An alternative modified score function, proposed by \citet{Kenne17}, aims at median bias reduction (median BR). Mean BR estimator has smaller mean bias than ML and equivariant under linear transformations of the parameters, whereas median BR estimator is componentwise third-order median unbiased in the continuous case and equivariant under componentwise monotone reperameterizations. We study the proposed adjusted score methods through a simulation study and an application, comparing their performance with respect to ML.

%%%%%%%%%%%%%%%%%%%%%%%%%%%%%%
\section{Dirichlet distribution}
%%%%%%%%%%%%%%%%%%%%%%%%%%%%%%
Let $y_i=(y_{i1}, \ldots, y_{im})^\top$, $i=1, \ldots, n$, be independent realizations of the $m$-dimensional Dirichlet random vectors parameterized by $\alpha=(\alpha_1, \ldots, \alpha_m)^\top$, with $\alpha_k>0$, $k=1, \ldots, m$.
%  and subject to $y_{ik}>0$, $k=1, \ldots, m$, and $\sum_{j=1}^{m}y_{ij}=1$. 
The probability density function of $Y_i\sim Dir(\alpha)$ is 
$$f_ {Y_i}(y_i;\alpha)=\frac{\Gamma(\sum_{j=1}^{m} \alpha_j)}{\prod_{j=1}^{m}{\Gamma(\alpha_j)}}\prod_{j=1}^{m} y_{ij}^{\alpha_j-1}$$
%where $\bold{B}(\alpha)=(\prod_{j=1}^{m}{\Gamma(\alpha_j)})/\Gamma(s)$, with $s=\sum_{j=1}^{m} \alpha_j$, is the multinomial beta function. 
with $y_{ik}>0$, $k=1, \ldots, m$, and $\sum_{j=1}^{m}y_{ij}=1$.
The log-likelihood is $$\ell(\alpha)=n\bigg \{\log \Gamma(s)-\sum_{j=1}^{m}\log \Gamma(\alpha_j)+\sum_{j=1}^{m}\alpha_j z_j\bigg \},$$ where $z_j=(\sum_{i=1}^{n}\log y_{ij})/n$. The log-likelihood is globally concave and the ML estimate needs to be obtained numerically.  Parameter estimation is usually carried out through a Fisher scoring-type algorithm with a sensible choice of the starting value.  \citet{Wicker}'s proposal seems to be  a stable initialisation. 

%%%%%%%%%%%%%%%%%%%%%%%%%%%%%%
\section{Modified score functions}
%%%%%%%%%%%%%%%%%%%%%%%%%%%%%%
For a general parametric  model with $m$-dimensional  parameter $\alpha$ and log-likelihood $\ell(\alpha)$,  based on a sample of size $n$, let $U_r=U_r(\alpha)=\partial \ell(\alpha) / \partial \alpha_r$ 
be the $r$-th component of the score function $U(\alpha)$, $r=1,\ldots,m$.
% and $-U_{rs}$ are the generic entries 
Let $j(\alpha)=-\partial^2 \ell(\alpha)/\partial \alpha\partial \alpha^\top$ be the observed information and  $i(\alpha)=E_\alpha\{j(\alpha)\}$ the expected information.

In order to reduce the bias of the ML estimator, \citet{Firth} proposes  a suitable modified score  aiming at   mean BR,  of the form
$$\tilde U(\alpha)=U(\alpha)+A^*(\alpha),$$ 
where the vector $A^*(\alpha)$ has components $A^*_r= \frac{1}{2} {\rm tr} \{i(\alpha)^{-1} [P_r+Q_r]\}$, with $P_r=E_\alpha\{U(\alpha)U(\alpha)^\top U_r\}$ and $Q_r=E_\alpha\{-j(\alpha) U_r\}$, $r=1,\ldots, m$.  The resulting estimator, $\hat\alpha^*$, has a mean bias of order $O(n^{-2})$, less than $O(n^{-1})$ of the ML estimator. 

A competitor estimator, $\tilde\alpha$, with accurate median centering property is obtained as solution of the estimating equation based on the modified score \citep{Kenne20}
$$\tilde U(\alpha)=U(\alpha)+\tilde A(\alpha),$$
with $\tilde A(\alpha)=A^*(\alpha)-i(\alpha)F(\alpha)$. 
The vector $F(\alpha)$ has  components $F_r=[i(\alpha)^{-1}]_r^\top\tilde F_r$, 
where $\tilde F_r$ has elements $\tilde F_{r,t}= {\rm tr}\{h_r[(1/3)P_t+(1/2)Q_t]\}$, $r,t=1,\ldots,m$, 
with the matrix $h_r$ obtained as $h_r=\{[i(\alpha)^{-1}]_r[i(\alpha)^{-1}]^\top_r\} /i^{rr}(\alpha)$, $r=1,\ldots,m$. Above, we denoted by $[i(\alpha)^{-1}]_r$ the $r$-th column of $i(\alpha)^{-1}$ and by $i^{rr}(\alpha)$ the  $(r,r)$ element of $i(\alpha)^{-1}$.

In the continuous case, each component of $\tilde \alpha$, $\tilde\alpha_r$, $r=1,\ldots,m$, is median unbiased with error of order $O(n^{-3/2})$,  i.e. Pr$_{\alpha}(\tilde\alpha_r\leq\alpha_r)=\frac{1}{2}+O(n^{-3/2})\,$, compared with  $O(n^{-1/2})$ of ML estimator. Both  $\hat\alpha^*$ and   $\tilde\alpha$ have the same asymptotic distribution    as that of the ML estimator, that is  $\hat \alpha\stackrel {\cdot}\sim \mathcal{N}_m(\alpha,i(\alpha)^{-1})$.

%%%%%%%%%%%%%%%%%%%%%%%%%%%%%%
\section{Bias reduction in Dirichlet regression models}
%%%%%%%%%%%%%%%%%%%%%%%%%%%%%%
Mean and median bias reduction have been extended to Dirichlet regression models. Following \citet{Maier}, we obtained the needed quantities for the adjusted score equations considering two parameterization of the Dirichlet's distribution, referred as common and alternative parameterization. Extensive simulation studies and applications will appear in a subsequent work. 

%%%%%%%%%%%%%%%%%%%%%%%%%%%%%%
\section{Simulation study}
%%%%%%%%%%%%%%%%%%%%%%%%%%%%%%
Through a simulation study, with small sample size settings, we compared the performance of the ML, mean and median BR estimators, $\hat \alpha$, 
$\hat \alpha^*$ and $\tilde \alpha$, respectively.
  The estimators are compared in terms of empirical probability of underestimation (PU), estimated relative mean bias (RB), and empirical coverage of the 95\% Wald-type confidence interval (WALD). The three performance measures are expressed in percentages.

\begin{table}[htbp!]\caption{Estimation of parameter  $\alpha=(\alpha_1, \alpha_2, \alpha_3)$. Simulation results  for ML ($\hat \alpha$), mean BR ($\hat \alpha^*$) and median BR ($\tilde \alpha$) estimators.}
\label{tab:1}
\centering
\begingroup
\renewcommand{\arraystretch}{1.0} % Default value: 1; sets the vertical (row) spacing
\resizebox{\textwidth}{!}{  
\begin{tabular}{p{0.5cm}p{0.5cm}p{1.2cm}p{1.2cm}p{1.2cm}p{1.2cm}p{1.2cm}p{1.2cm}p{1.2cm}p{1.2cm}p{1.2cm}}
 \noalign{\smallskip}  \cmidrule{1-11} 
& &    \multicolumn{3}{|c|}{ $n=10$ }& \multicolumn{3}{|c|}{$n=20$} & \multicolumn{3}{|c}{$n=40$} \\ 
\noalign{\smallskip}  \cmidrule{1-11} 
 &$\alpha$& \multicolumn{1}{|l}{PU} & RB & WALD &  \multicolumn{1}{|l}{PU} & RB & WALD & \multicolumn{1}{|l}{PU}  & RB  &  WALD \\ 
\noalign{\smallskip}  \cmidrule{1-11} 
\multirow{9}{*}{S1}  
&$\hat\alpha_1$ & 40.89 & 20.89 & 96.34 & 43.19 & 9.23 & 95.69 & 44.40 & 4.39 & 95.63 \\ 
&$\hat\alpha^*_1$ & 60.87 & -0.17 & 90.25 & 56.75 & 0.01 & 92.75 & 54.30 & 0.05 & 94.09 \\ 
&$\tilde \alpha_1$ & 50.26 & 10.39 & 94.31 & 49.54 & 4.69 & 94.75 & 49.11 & 2.27 & 95.04 \\ \cmidrule{2-11} 
&$\hat\alpha_2$& 40.77 & 21.08 & 96.12 & 43.21 & 9.39 & 95.79 & 45.16 & 4.48 & 95.48 \\ 
 &$\hat\alpha^*_2$& 60.32 & -0.03 & 89.67 & 57.29 & 0.16 & 92.92 & 55.09 & 0.13 & 94.11 \\ 
&$\tilde \alpha_2$ & 50.04 & 10.56 & 94.07 & 49.84 & 4.84 & 94.76 & 49.96 & 2.35 & 95.03 \\ \cmidrule{2-11} 
&$\hat\alpha_3$& 39.93 & 21.13 & 96.54 & 43.40 & 9.24 & 95.82 & 45.32 & 4.50 & 95.19 \\ 
&$\hat\alpha^*_3$& 60.55 & 0.02 & 90.35 & 57.71 & 0.02 & 92.97 & 54.87 & 0.15 & 93.84 \\ 
&$\tilde \alpha_3$& 49.50 & 10.61 & 94.36 & 50.19 & 4.70 & 94.67 & 49.97 & 2.37 & 94.64 \\ 
\noalign{\smallskip}  \cmidrule{1-11} 
\multirow{9}{*}{S2}  
&$\hat\alpha_1$ & 38.22 & 33.48 & 96.57 & 40.27 & 14.68 & 96.11 & 44.13 & 6.70 & 95.84 \\ 
&$\hat\alpha^*_1$& 63.91 & -0.61 & 86.97 & 58.66 & 0.40 & 91.61 & 56.60 & 0.15 & 93.70 \\ 
&$\tilde \alpha_1$& 49.94 & 16.12 & 93.30 & 49.16 & 7.51 & 94.53 & 50.24 & 3.43 & 95.11 \\\cmidrule{2-11} 
&$\hat\alpha_2$& 40.40 & 23.22 & 96.23 & 42.71 & 10.15 & 95.88 & 44.03 & 4.92 & 95.23 \\ 
 &$\hat\alpha^*_2$& 61.35 & -0.08 & 89.16 & 57.35 & 0.13 & 92.94 & 54.38 & 0.22 & 93.90 \\ 
&$\tilde \alpha_2$& 50.20 & 11.27 & 93.73 & 50.24 & 5.04 & 95.08 & 49.34 & 2.54 & 94.77 \\ \cmidrule{2-11} 
&$\hat\alpha_3$ & 42.84 & 15.08 & 96.01 & 45.15 & 6.84 & 95.46 & 46.63 & 3.23 & 95.51 \\ 
& $\hat\alpha^*_3$ & 59.75 & -0.04 & 91.10 & 56.75 & 0.02 & 93.12 & 54.26 & -0.02 & 94.26 \\ 
&$\tilde \alpha_3$& 49.77 & 8.26 & 94.54 & 50.02 & 3.80 & 94.81 & 49.99 & 1.79 & 95.23 \\ 
\noalign{\smallskip}\cmidrule{1-11} 
\multirow{9}{*}{S3}  
&$\hat\alpha_1$& 33.06 & 26.14 & 96.03 & 38.48 & 11.28 & 95.47 & 42.29 & 5.37 & 95.40 \\ 
&$\hat\alpha^*_1$& 59.07 & 0.25 & 89.37 & 56.72 & -0.14 & 92.14 & 54.32 & -0.03 & 93.67 \\ 
&$\tilde \alpha_1$& 49.75 & 9.06 & 92.88 & 50.12 & 3.73 & 93.95 & 50.01 & 1.80 & 94.61 \\ \cmidrule{2-11} 
&$\hat\alpha_2$& 33.88 & 25.49 & 95.79 & 38.46 & 11.05 & 95.62 & 42.69 & 5.26 & 95.29 \\  
&$\hat\alpha^*_2$ & 58.98 & 0.16 & 89.29 & 56.15 & -0.13 & 92.31 & 54.24 & -0.02 & 93.52 \\ 
&$\tilde \alpha_2$& 50.28 & 8.91 & 93.13 & 49.98 & 3.73 & 94.15 & 50.21 & 1.80 & 94.49 \\\cmidrule{2-11} 
&$\hat\alpha_3$& 35.06 & 23.68 & 96.05 & 39.47 & 10.19 & 95.58 & 42.96 & 4.79 & 95.32 \\ 
&$\hat\alpha^*_3$& 58.61 & 0.26 & 89.79 & 56.26 & -0.13 & 92.39 & 54.55 & -0.10 & 93.90 \\ 
&$\tilde \alpha_3$& 49.31 & 8.81 & 93.52 & 49.96 & 3.66 & 94.38 & 50.02 & 1.70 & 94.50 \\ 
\noalign{\smallskip}\cmidrule{1-11} 
\multirow{9}{*}{S4}  
&$\hat\alpha_1$& 33.22 & 25.32 & 96.32 & 38.12 & 10.92 & 95.54 & 41.66 & 5.19 & 95.69 \\ 
&$\hat\alpha^*_1$& 58.13 & 0.32 & 89.37 & 56.70 & -0.12 & 92.27 & 53.96 & -0.04 & 94.04 \\ 
&$\tilde \alpha_1$& 49.43 & 8.78 & 93.34 & 50.34 & 3.61 & 94.06 & 49.75 & 1.73 & 94.70 \\\cmidrule{2-11} 
&$\hat\alpha_2$& 33.26 & 25.32 & 96.34 & 38.43 & 10.98 & 95.34 & 41.50 & 5.18 & 95.17 \\ 
&$\hat\alpha^*_2$& 58.25 & 0.32 & 89.46 & 56.33 & -0.07 & 92.35 & 54.77 & -0.05 & 93.81 \\
& $\tilde \alpha_2$& 49.16 & 8.78 & 93.31 & 50.15 & 3.67 & 94.08 & 50.21 & 1.72 & 94.59 \\ \cmidrule{2-11} 
&$\hat\alpha_3$& 33.25 & 25.45 & 96.31 & 38.62 & 10.98 & 95.64 & 41.91 & 5.18 & 95.36 \\ 
&$\hat\alpha^*_3$& 58.65 & 0.43 & 89.55 & 56.35 & -0.07 & 92.65 & 54.85 & -0.05 & 94.01 \\ 
&$\tilde \alpha_3$& 49.00 & 8.90 & 93.21 & 50.14 & 3.67 & 94.27 & 50.09 & 1.71 & 94.71 \\ 
\noalign{\smallskip}\cmidrule{1-11} 
\end{tabular}}
\endgroup
\end{table}

We consider the sample sizes $n=10, 20, 40$, and, for each of 10000 replications, we draw  samples of  independent observations from 3-dimensional Dirichlet random vector, with true parameter value $\alpha_0$. Combination of small and large true parameter values with equal and different values are considered. In particular, we perform the  study under the settings  $\alpha_0=(0.25,0.25,0.25)$  (S1), $\alpha_0=(0.6,0.3,0.1)$ (S2), $\alpha_0=(12,6,2)$ (S3), and $\alpha_0=(40/3,40/3,40,3)$ (S4).

Table 1 shows the numerical results of the simulations.  For all settings, mean and median BR estimators proved to be remarkably accurate in achieving  their own goals, respectively, and are preferable to ML estimators. The poor coverage of  the mean BR estimator is implied by the strong shrinkage effect of the estimator, whereas median BR shows empirical coverage  closer to nominal values. The good performances of the ML estimator in terms of empirical coverages, especially when compared with mean BR, are overwhelmed by very large estimated relative mean bias and a noteworthy overestimation of the true parameter. 

\section{Application}
We consider the serum-protein data of Pekin-ducklings analysed in \citet{Ng}, coming from \citet{Mosimann}. 
Data concerns blood serum proportions of $n=23$ sets of Pekin-ducklings, characterized by having the same diet in each set. For the $i$-th set, $i=1, \ldots, 23$,  the proportion of pre-albumin ($y_{i1}$), albumin ($y_{i2}$) and globulin ($y_{i3}$), are reported.  
%For $n=23$ sets of Pekin-ducklings, with the same diet in each set, blood serum proportions (pre-albumin, albumin, globulin) are reported.
Ternary plot, in Figure 1,  shows in two-dimensions the distibution of $y_i=(y_{i1}, y_{i2}, y_{i3})^\top$ on the simplex. Data shows that for a small amount  of pre-albumina there is  about a  50/50 composition of albumin and globulin.
\begin{figure}[htbp]
\centering{
\includegraphics[scale=0.39085] {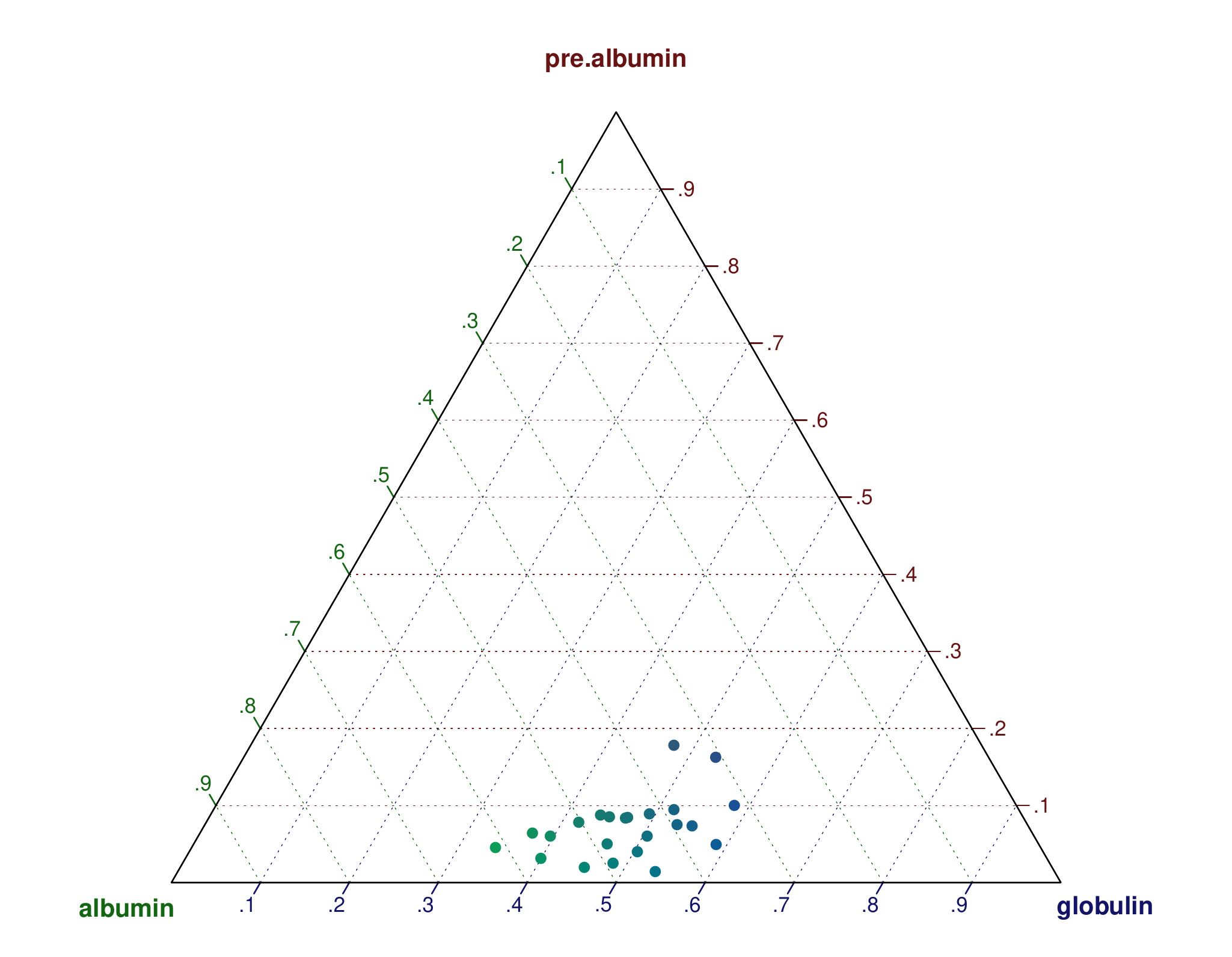}}
\caption{Serum-protein data of Pekin-ducklings. Ternary plot.}
\label{fig:2}       % Give a unique label
\end{figure}

  Table 2  reports point and interval estimates of the  parameters, by using ML, mean and median BR. It is noteworthy the shrinkage effect of the mean BR estimator.  Median BR estimates are intermediate between those of  mean BR and  ML estimates, as well as for the estimated standard errors. As a result of the shrinkage effect of the mean and median BR estimators, the 95\% Wald-type confidence intervals for mean BR and median BR are narrower than those of ML.

\label{sec:6}
\begin{table}[ht] \caption{Serum-protein data of Pekin-ducklings. Estimates of parameter $\alpha=(\alpha_1, \alpha_2, \alpha_3)$, estimated standard errors and 95\% Wald-type confidence intervals (95\% Wald CI) using ML,  mean  and median BR.}
\centering
\resizebox{\textwidth}{!}{
 \begin{tabular}{p{1cm}p{2.0cm}p{3cm}p{2.5cm}}
 \noalign{\smallskip}  \cmidrule{1-4} 
 $\alpha$  & Estimate & Standard error & 95\% Wald CI \\ 
 \noalign{\smallskip}  \cmidrule{1-4} 
$\hat \alpha_1$   & 3.22 & 0.68 & 1.89 - 4.54 \\ 
$\hat \alpha^*_1$    & 2.95 & 0.62 & 1.73 - 4.17 \\ 
$\tilde \alpha_1$      & 3.04 & 0.64 & 1.79 - 4.30 \\
 \noalign{\smallskip}  \cmidrule{1-4} 
$\hat \alpha_2$& 20.38 & 4.32 & 11.91 - 28.86 \\ 
$\hat \alpha^*_2$& 18.59 & 3.95 & 10.84 - 26.33 \\ 
$\tilde \alpha_2$  & 19.19 & 4.08 & 11.20 - 27.18 \\
 \noalign{\smallskip}  \cmidrule{1-4} 
$\hat \alpha_3$& 21.69 & 4.60 & 12.67 - 30.70 \\ 
$\hat \alpha^*_3$  & 19.77 & 4.20 & 11.54 - 28.01 \\ 
$\tilde\alpha_3$  & 20.41 & 4.34 & 11.92 - 28.91 \\ 
 \noalign{\smallskip}  \cmidrule{1-4} 
\end{tabular}}
\end{table}

\end{document}